\documentclass[aps,physrev,preprint,groupedaddress]{revtex4-2}
\usepackage{lineno}
\usepackage{comment}
\usepackage{amsmath}
\usepackage{graphicx}
\usepackage{color}

\begin{document}


\title{Digitally enhanced Multi-wavelength Stabilization using a Passive Fiber Frequency Reference}



\author{Hilma Karlsson}
\email{hilmak@kth.se}
\author{Vaishali Adya}
\affiliation{Department of Applied Physics, KTH Royal Institute of Technology, Roslagstullsbacken 21, Stockholm SE-106 91, Sweden}

\author{Robert L. Ward}
\author{Chathura P. Bandutunga}
\email{chathura.bandutunga@anu.edu.au}
\affiliation{Centre for Gravitational Astrophysics, College of Science, Australian National University, Acton, Australian Capital Territory 2601, Australia}%



\date{\today}

\begin{abstract}
High-stability frequency stabilization between lasers is essential for applications such as precision metrology, frequency dissemination, and quantum communications. We demonstrate the use of code-based multiplexing of a passive fiber-based interferometer for frequency stabilization and transfer between lasers separated by 85 GHz in frequency using a single shared passive reference. Out-of-loop characterization places an upper bound of sub-kHz/$\sqrt{\text{Hz}}$ on the transferred frequency noise at Fourier frequencies above $40$~mHz, with a differential fractional stability of $7\times10^{-13}$  at 1 second. We also demonstrate continuous laser tunability and characterize the residual length noise coupling. Finally, we provide a framework for scaling the architecture to higher laser counts, offering a pathway to scalable, frequency agile stabilization of multiple light sources without the overhead of an optical frequency comb.
\end{abstract}

\maketitle

\section{Introduction}
Relative phase and frequency stability between optical carriers is a key requirement in precision photonics, coherent communications, and quantum networks. Quantum key distribution (QKD) protocols, such as continuous-variable QKD (CV-QKD) and twin-field QKD (TF-QKD), are particularly sensitive to phase stability in the generation, transmission, and detection of quantum states. In CV-QKD, the efficacy of the protocol relies on establishing or recovering a stable phase relation between the received signal and the local oscillator. In practical implementations that are based on a local local oscillator, self-referenced phase recovery, or pilot-assisted reconstruction of quantum states, differential phase and frequency noise contributes directly to phase estimation error and excess noise \cite{huang2015nolo,qi2015llo,soh2015selfref,laudenbach2019pilot}. 
In TF-QKD, long distance operation relies on achieving high visibility interference at the central measurement station, despite two optical fields being generated by remote users separated by up to 1000~km. This places stringent requirements on optical frequency matching, and relative phase stability \cite{lucamarini2018tfqkd,clivati2022twinfield,liu2023tfqkd1000km}. In implementations based on dual-band stabilization, an auxiliary wavelength co-propagates with the quantum signals, typically in an adjacent ITU wavelength channel, in order to track and suppress large-amplitude phase drift introduced by long-distance fiber transmission. For the auxiliary wavelength to correctly track the phase drifts of the primary wavelength of the quantum signals, the auxiliary wavelength needs to be wavelength referenced back to the primary wavelength, prior to link injection. This necessitates the local transfer of frequency stability when preparing the auxiliary wavelength, with multi-GHz separations.

Passive fiber interferometers based on path imbalanced Mach Zehnder and Michelson interferometers provide a promising route to the stabilization required for such experiments. These interferometers convert laser frequency fluctuations into measurable phase fluctuations using standard fiber components, and are therefore attractive as passive frequency discriminators \cite{kefelian2009delay,kong2015fiberinterferometers,jeon2023palm-sized-559}. The achievable performance of these references is fundamentally limited by thermodynamic path length fluctuations, environmental perturbations, and parasitic interference processes \cite{wanser1992phasenoise,ngo2014,zhang2021double-f5c,jeon20251015-level-ac3}. The locking process itself also contributes, with discriminator nonlinearity coupling out of band phase noise contributing to a white noise floor that is suppressed only at high loop bandwidth \cite{grabielle2025locking-bb8}. For full multi-wavelength operation, an additional challenge is that common mode cancellation of interferometer length noise is not exact when the interrogating lasers are separated in wavelength \cite{luiten2020, show2025thesis}. 

In previous demonstrations of multi-wavelength frequency transfer methods for fiber references, a bi-directional interferometer architecture was used with an active modulator within to create two self-heterodyne measurements \cite{luiten2020, show2025thesis}. These demonstrations showed a 1 Hz/$\sqrt{\text{Hz}}$ relative stability from 80 Hz to 800 Hz for lasers separated by 666 nm with a 100 kHz control bandwidth and below 1 Hz/$\sqrt{\text{Hz}}$ transfer noise from 4 Hz to 30 kHz for a frequency separation of 400 MHz respectively 18 nm, with a locking bandwidth of around 280 kHz. A demonstration by Grabrielle et al. uses two 1542 nm lasers with a separation of 450 MHz, stabilized using frequency non-degenerate Pound-Drever-Hall locking on a Michelson interferometer with an operational frequency band between 120 Hz and 8 kHz, with discrete lock points spaced at half an interferometer free spectral range\cite{grabielle2025locking-bb8}.

In this work, we combine multi-frequency transfer methods on fiber references with digitally enhanced homodyne interferometry (DEHoI)\cite{shaddock2007dehi,mcrae2014deofr,ngo2014,zhang2021double-f5c} techniques for frequency stabilization. In our system, two fiber lasers are phase modulated with independently delayed quadrature phase shift keying (QPSK) codes outside the interferometer and therefore interrogate a fully passive path imbalanced, uni-directional Mach Zehnder architecture. Using code division multiplexed digital decoding, independent complex interferometric readouts are recovered for each wavelength channel from a single photodetector. In our case the difference between these two measurements is used to stabilize the two lasers to each other. This architecture simultaneously absolves the need for active components within the fiber reference, its unidirectionality makes it immune to first order fiber backscattering, the digital readout retains the full continuous fringe tracking and acquisition flexibility of a fiber interferometer topology and it is extensible to higher laser counts using the one fiber reference and detection chain. 

We demonstrate optical frequency stabilization between two fiber lasers separated by 85.42~GHz. The out-of-loop differential measurement shows an Allan deviation of $7\times10^{-13}$ at 1~s, continuous operation over 3.5~days with the transferred laser bounded within $\pm 300$~kHz with a 100Hz control bandwidth, and digitally controlled lock-point tunability with 122 Hz resolution. Finally, we characterize the residual interferometer length noise coupling at hour timescales, measuring wavelength dependent cross-coupling at the 700 ppm level and thereby identifying the dominant limitation to medium and long term stability.

This initial demonstration lays the groundwork for a flexible, scalable, and frequency agile platform for referencing and transferring frequency stability between multiple lasers of common or differing wavelengths. While our motivation is driven by quantum secure communications, we can see the technique can be broadly relevant for multi-wavelength ranging \cite{DeGroot1991}, tunable terahertz generation \cite{he2025highly-419,danion2014dual-f3d, kittlaus2021low-noise-558} and optical frequency transfer \cite{ebenhag2011active-983,Roslund2025}. Section~II describes the multiplexed frequency readout, Sec.~III the experimental implementation, Sec.~IV the stability measurements, Sec.~V the cross-coupling measurements, Sec.~VI the tunability demonstration, Sec.~VII discusses scaling requirements, and Sec.~VIII concludes the paper.

\section{Multiplexed Frequency Readout}
Optical fiber references have been extensively demonstrated using several techniques, notably delay heterodyne and baseband readout methods. More recently, there have been demonstrations using DEHoI on fiber references, enabling a passive architecture capable of stability down to $0.1\; \text{Hz}/\sqrt{\text{Hz}}$ \cite{bandutunga2020infrasonic-37a,zhang2021double-f5c,grabielle2025locking-bb8}.

DEHoI operates by translating spread-spectrum communication techniques into the optical domain. In a DEHoI-based frequency reference, an electro-optic phase modulator (EOM) applies a pseudo-random noise (PRN) code, $c(t)$, to the interrogation laser prior to the interferometer. This modulation phase, $\phi(t)$, spreads the optical carrier across a wide bandwidth. We can represent this as an optical carrier with a code modulated phase:
\begin{align}
E_1(t) = Ae^{i\omega t + ic(t)}
\label{eq:efield}
\end{align}
where $c(t)$ represents the code at delay of zero and $\omega$ is the optical frequency for Laser 1. Nominally the modulation depth of the code is such that $c(t)$ is composed of four levels at multiples of $\pi/4$ radians, forming a quadrature phase shift key. This is important later as each of these steps maps to a different quadrature in the complex IQ basis.

After propagating Eqn. \ref{eq:efield} through a Mach Zehnder interferometer with delays of $\tau_S$ and $\tau_L$ for the short and long arms of the interferometer respectively, the power detected at balanced homodyne detection can be represented as per the standard two beam interference expression in Eqn. \ref{eq:pd}:
\begin{align}
    P(t) &= \alpha A_1^2\cos[c(t - \tau_S) - c(t - \tau_L) + \Delta\phi_1(t)],
    \label{eq:pd}
\end{align}
where $c(t - \tau_S) - c(t - \tau_L)$ is the code phase difference, and $\Delta\phi(t)$ is the interferometer phase difference measured by Laser 1. The interferometer loss is given by $\alpha$. In an arm-length mismatched interferometer, $\Delta\phi_1(t)$ is proportional to the laser frequency noise $\Delta\nu_1(t)$ scaled by the interferometer free-spectral range, thus providing our measurement of laser frequency with respect to the interferometer. 

An important property of DEHoI we utilize is the ability to perform phase reconstruction using upfront (common) modulation. This process, called double demodulation, reconstructs the phase difference between the modulation present on both arms of the interferometer \cite{Sutton2012DigitallyInterferometry}. As the repetition length of the code is significantly longer than either arm of the interferometer, the differential modulation between arms is unique, and unambiguous in terms of total delay through the system. Functionally, it means that $c(t - \tau_S)$ has minimal correlation with $c(t - \tau_L)$, and are thus considered distinct, non-degenerate modulation over the interferometer timescale. 

IQ demodulation for this well defined differential code with a known system propagation delay recovers the interferometer phasor, which can then be used to compute the interferometric phase. This absolves the system from requiring active components within the interferometer, and is what enables the system to multiplex the frequency noise measurements of multiple lasers simultaneously. 

To demodulate this signal, we perform effectively an IQ demodulation, as routine with heterodyne detection. Here though, instead of a heterodyne frequency, we have a spread-spectrum encoded signal we are demodulating against.
In digital signal processing (DSP), we therefore multiply the resultant signal by a time-delayed replica of the measured codes, $c(t- \tau_S) - c(t - \tau_L)$, projected into the complex IQ basis, and time averaged over the duration of a full code length, given by $\tau_{\text{code}}$. This process is depicted in schematic form in Fig. \ref{fig:dsp} b), and we can write these projections as per Eqn. \ref{eq:IQdemodulation1} and \ref{eq:IQdemodulation2}:
\begin{align}
    I_1 = \sum_{t = 0}^{\tau_{\text{code}}}\cos(c(t - \tau_S) - c(t - \tau_L))\times P(t)\label{eq:IQdemodulation1}\\
    Q_1 = \sum_{t = 0}^{\tau_{\text{code}}}\sin(c(t - \tau_S) - c(t - \tau_L))\times P(t)
    \label{eq:IQdemodulation2}
\end{align}

This correlation process acts as a synthetic range gate, isolating the interference signal associated with a specific optical times-of-flight, $\tau_S$ - $\tau_L$. By computing the autocorrelation of the code during the demodulation process, the technique effectively rejects interferometric signals from shorter or longer paths, and has been demonstrated to suppress effects from Rayleigh backscatter and multi-path interference \cite{zhang2019multi-target-68a,Ngo2016SuppressingInterferometers,dey2021algebraic-04f}. This is akin to a heterodyne detection isolating frequencies outside its filter passband frequency.

\begin{figure}
    \centering
    \includegraphics[width=0.8\linewidth]{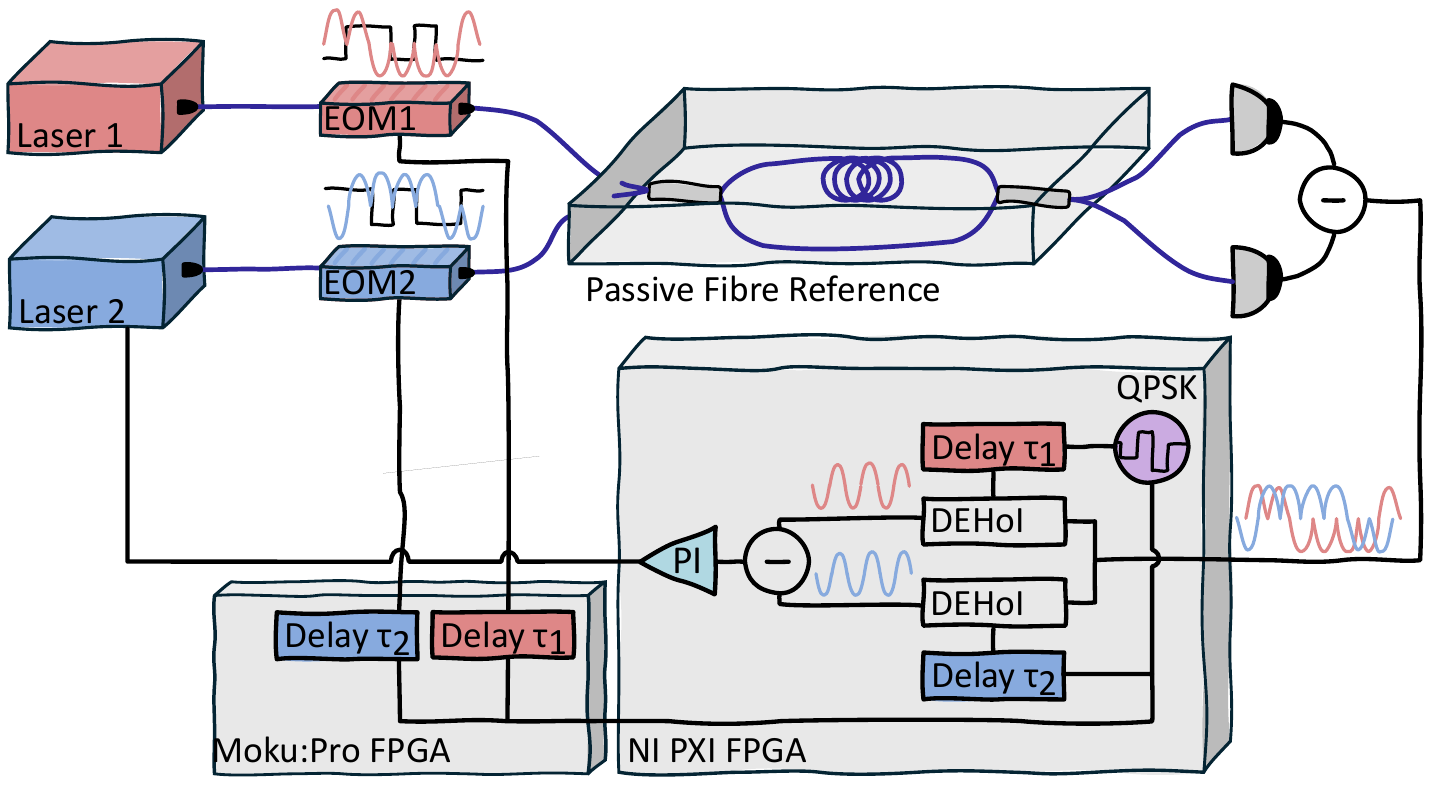}
    \caption{Schematic Diagram for digital interferometry (DI) decoding of multi-wavelength system. Each laser is phase modulated with a delayed QPSK code using an EOM and then fed into a passive fiber reference. The multi-interferometric readout from the passive fiber reference is demodulated with a delayed copy of the unique QPSK code, yielding each laser's phase measurement.}
    \label{fig:DI_decoding}
\end{figure}

In this work, we exploit this range-gating and double demodulation properties of DEHoI by encoding two lasers each with a uniquely delayed DEHoI modulation outside of a passive, delay line Mach Zehnder interferometer, as depicted in Fig. \ref{fig:DI_decoding}. At photodetection, this effectively reads as two interferometers, each with a unique offset to their transit delay, $\tau_i$. 

The addition of second laser adds another homodyne interference term to the photodetector readout. This yields Eqn. \ref{eq:pd_dual}, which shows the self-interference (homodyne) terms in the system. 
\begin{align}
    P(t) &= \alpha A_1^2\cos[c(t - \tau_S - \tau_1) - c(t - \tau_L- \tau_1) + \Delta\phi_1(t)] \nonumber \\ &\quad + \alpha A_2^2\cos[c(t - \tau_S- \tau_2) - c(t - \tau_L- \tau_2) + \Delta\phi_2(t)]
    \label{eq:pd_dual}
\end{align}
In Eqn. \ref{eq:pd_dual}, we see that the two interferometers have two unique code pairs, made possible by combinations of the four unique code delays $\tau_1$, $\tau_2$, $\tau_S$ and $\tau_L$. In DSP, we can therefore demodulate for each interferometer phase in parallel, distinguishing between the two interference terms by their identifying delays $\tau_1$ and $\tau_2$. Note that the heterodyne interference terms in this experiment have a beat frequency outside photodetection range, and they are also encoded with different code combinations to those in Eqn. \ref{eq:pd_dual}. Thus they are both code and frequency separable.

In the experiment, we demonstrate the identification, isolation and demodulation of each laser by this method, yielding independent measurements from the same fiber frequency reference. The phase subtraction between the two measurements then yields an error signal which has been demonstrated to be first order transparent to interferometer length noise \cite{luiten2020}. This phase subtraction can be computed as the difference between two phasors. We can compute the phasor difference by performing a complex multiplication between phasor $\bar{A}_1$ and the conjugate of phasor $\bar{A}_2$ \cite{dey_common-mode_2024}:
\begin{align}
\label{eq:diffphase}
    \Delta\phi_{12} = \angle \bar{A}_1.\bar{A}_2^* &= \arctan\left(\frac{I_1 Q_2 - Q_1 I_2}{I_1I_2 + Q_1Q_2}\right)
\end{align}
We close the loop by stabilizing one laser to the second using the $\Delta Q$ component as an error signal. For small angles, the error signal $1 Q_2 - Q_1 I_2 $ is proportional to $ \Delta\phi_{12} $. This is valid while the system remains locked, with an additional loop gain factor from the amplitude of the two phasors.

\section{Experimental Implementation}
\begin{figure}
    \centering
    \includegraphics[width=0.8\linewidth]{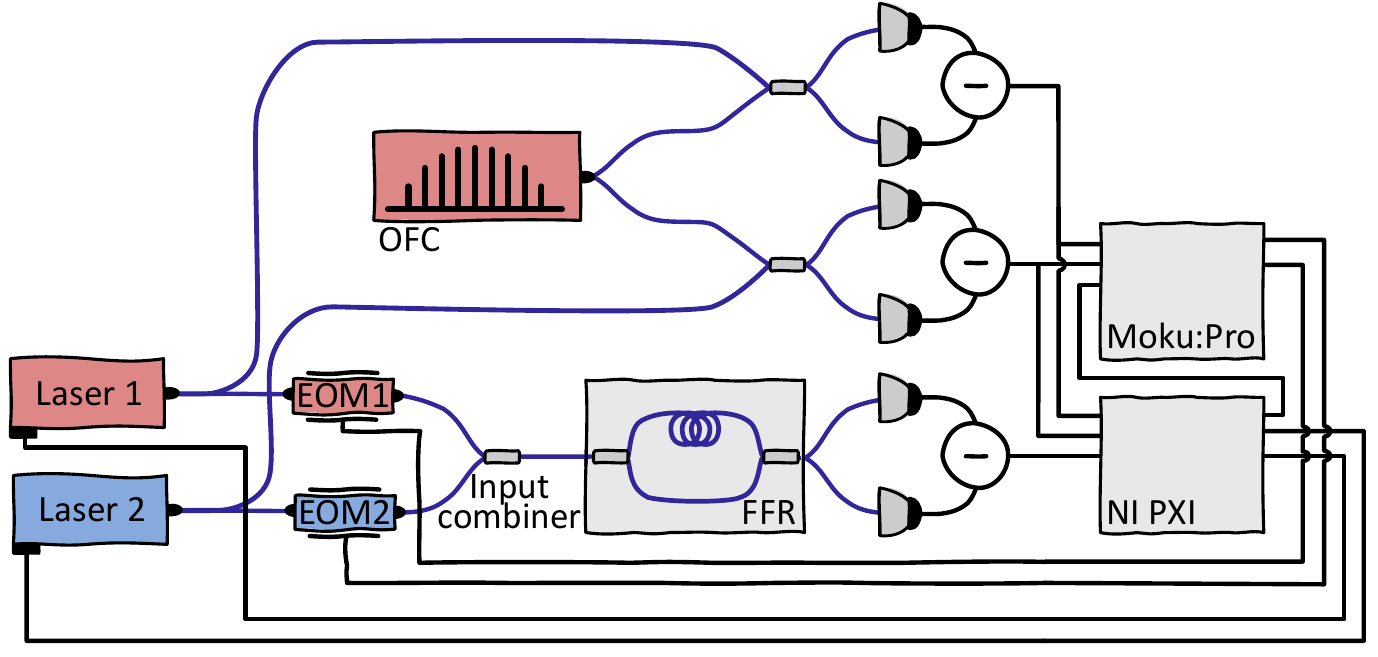}
    \caption{Optical hardware diagram. The optical setup consists of the two lasers under test that are split between a frequency stabilization system consisting of the arm-length unbalanced fiber interferometer, and a characterization system, consisting of two beat detection units with an RF referenced optical frequency comb.}
    \label{fig:experimentaldiagram}
\end{figure}

The experimental implementation consists of a two laser system interrogating a path imbalanced Mach Zehnder interferometer. We use two Koheras E15 fiber lasers as our lasers under test, with wavelengths set to 1549.9185 and 1549.2343 nm respectively. This provides a separation of 684.2 pm in wavelength, equivalently 85.42 GHz in frequency. Each laser is split between the fiber interferometer and an optical frequency comb which is used for characterization, as illustrated in Fig. \ref{fig:experimentaldiagram}.

\subsection{Frequency Stabilization System}
The path leading to the fiber interferometer is phase modulated using an Exail MPX-LN-01 electro-optic modulator (EOM). The modulation is a QPSK code with a symbol frequency of 50 MHz. At a modulation depth of $\pi/4$, this yields a 4 point phase constellation with phase steps at $-3\pi/4$, $-\pi/4$, $\pi/4$ and $3\pi/4$. The second laser is also modulated with a QPSK code, which is delayed relative to the first by 1.2 microseconds. This is longer than the transit time for the field through the interferometer of $\sim$ 0.14 microseconds. Both the first and second modulated fields are then combined using a PLC 4x1 fiber combiner and enter the interferometer. The use of the 4x1 fiber combiner is to enable additional lasers to be added to the system.

The fiber interferometer is constructed using two 3 dB splitter/combiners and a 25 m delay-line coil in one arm of the interferometer. The code symbol frequency of 50 MHz corresponds to an equivalent symbol length of $\sim$ 4 m. This leads to 6.25 symbols of differential delay within the interferometer. The two outputs of the interferometer are detected using an Insight BPD-400 balanced photodetector and sent to digitization.

\begin{figure}
    \centering
    \includegraphics[width=\linewidth]{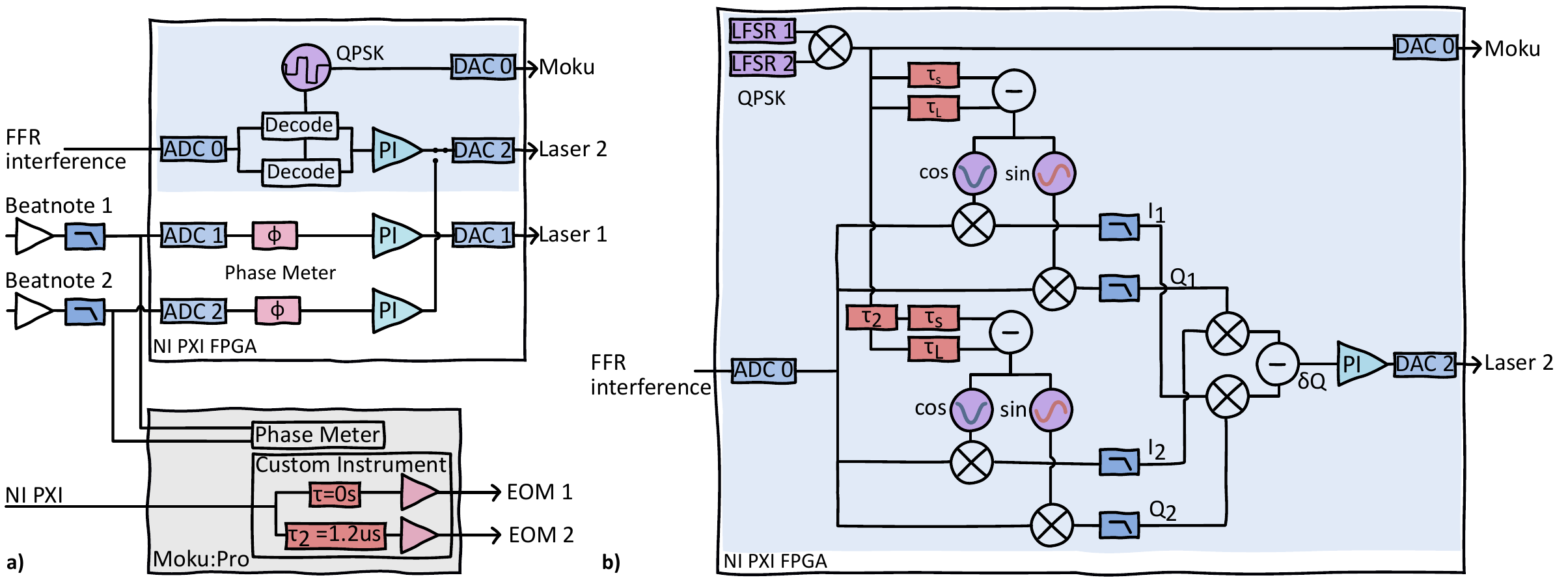}
    \caption{Digital signal processing block diagram. a) Shows the layout of the signal processing deployed on both a National Instruments PXI FPGA and a Moku:Pro. The blue shaded region in a) is expanded in b), showing a more detailed sketch of the demodulation of the frequency readout from the interferometer and frequency control systems for transferring frequency stability from one laser to the other.}
    \label{fig:dsp}
\end{figure}

The signal processing for the experiment is primarily deployed on a National Instruments PXI FPGA platform. This consists of a Xilinx Kintex-7 410T card (NI 7975) and an IF transceiver (NI 5783) which provides 4 analog inputs at 100 MS/s and 4 analog outputs at 200 MS/s. The NI platform has three main functions: generation of the QPSK modulation, demodulation of the frequency readout from the interferometer and frequency control systems for both lasers. A detailed map of the digital signal processing architecture is shown in Fig. \ref{fig:dsp}.

First, the generation of the QPSK modulation is done using a dual linear feedback shift register (LFSR) m-sequence generator. This generates two m-sequences delayed with respect to each other which are then logically combined to form the QPSK modulation sequence. Details of QPSK code generation for DEHoI are described in other works \cite{Sutton2012DigitallyInterferometry,ngo2014,Dey:25}. We use an 11-bit code length ($L_c$), yielding 2047 symbols per code and at the 50 MHz symbol rate, this results in a code update rate of 24.45 kHz.

The 4-level QPSK modulation is output from DAC0 on the transceiver and sent to a Moku:Pro platform from Liquid Instruments for distribution and amplification using a custom instrument. The custom instrument acts as a distribution amplifier with individual delay control, enabling a single QPSK code to be amplified and delayed uniquely across the 4 output channels of the Moku:Pro. In this demonstration, we use two of the output channels, with output 2 delayed relative to output 1 by 1.2 microseconds. Each output directly drives its corresponding laser's EOM.

The NI platform receives and digitizes the interferometer photodetector signal on ADC0. The digitized photodetector output is decoded based on the time-of-flight for each field through the short and long interferometer arms, $\tau_S$ and $\tau_L$. The differential phase for the interferometer can then be recovered. As each laser is modulated with a unique code, a single demodulation channel is able to isolate and recover the interferometer phase for a specific laser. Parallelizing the decoding process can then recover each laser's interferometer phase independently. This output is transferred to the frequency control system as an IQ data pair for each laser channel.

The frequency stabilization control system computes the differential phasor between the two IQ pairs using a complex multiplier and uses the quadrature component of the differential phasor for locking. A digital proportional-integral (PI) controller is used to implement the feedback control to actuate back onto laser 2 and stabilize it relative to laser 1. For the measurements presented here, the lock was operated with a control unity gain bandwidth of approximately 100 Hz. The actuation signal is output from DAC2 to the piezo control on laser 2.

\subsection{Characterization System}
Alongside the frequency stabilization system, an independent (out-of-loop) characterization system is implemented. We follow the other output from the initial splitter/coupler for each laser, which is directed to two beat detection units with a Menlo Systems optical frequency comb. Each beat detection unit consists of a 3 dB splitter/combiner followed by an Insight BPD-400 balanced photodetector. The output from each detector is amplified by a Minicircuits ZFL-500HLN preamplifier and filtered with a 62 MHz low pass filter. The output from the filter-amplifier stage is split between the Moku:Pro input channels 2 and 3, and NI PXI platform ADC2 and ADC3.

On the NI PXI platform, the digitized optical frequency comb beat notes with the lasers under test are measured using high-speed PLLs running with an update rate of 25 MHz to enable tracking of high-speed phase dynamics. The frequency output from each is optionally passed to individual PI-controllers which can offset frequency lock each laser to the frequency comb as required for specific measurements. In the case of laser 2, this PI-controller switches in-place of the frequency stabilization PI-controller when activated.

On the Moku:Pro side, the frequency comb measurements are tracked using the Liquid Instruments Phasemeter instrument. The instrument is set to 1 MHz bandwidth, again to capture the high frequency jitter of the optical frequency comb itself. 

The optical frequency comb has a repetition rate of 250 MHz, and is self-referenced to an Endrun Solutions Tycho II reference clock, with holdover provided by a ultra-stable crystal oscillator (US-OCXO). The same clock is used as the reference oscillator for the FPGA and Moku:Pro systems.

\section{Stability Measurements}
Characterizing the performance of the frequency stabilization system was done using an extended time-domain measurement. This was recorded with the Moku:Pro phasemeter instrument with a 1 MHz loop bandwidth at a decimated rate 37 S/s. The measurement was carried out over a 3.5 day period and both the characterization system and the frequency stabilization system remained stable over that duration. 

\begin{figure}
    \centering
    \includegraphics[width=0.7\linewidth]{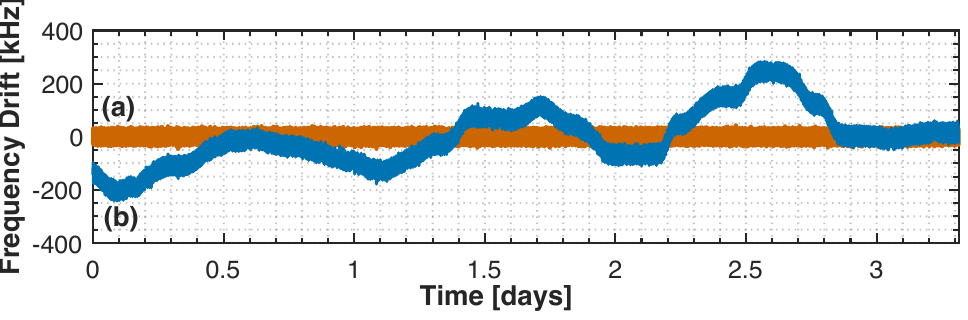}
    \caption{Time domain measurement of the out-of-loop beat notes of the two lasers with respect to the optical frequency comb. Trace (a) corresponds to laser 1 which is locked to the comb. Trace (b) corresponds to laser 2, which is offset locked to laser 1 via the fiber reference.}
    \label{fig:relativestability}
\end{figure}

In Fig. \ref{fig:relativestability} we plot the time series of both lasers. We see Laser 1 (a) was stable to the optical frequency comb for the duration of the measurement. Laser 2 (b) was then bounded in error to within $\pm 300$ kHz over the duration.

\begin{figure}
    \centering
    \includegraphics[width=0.95\linewidth]{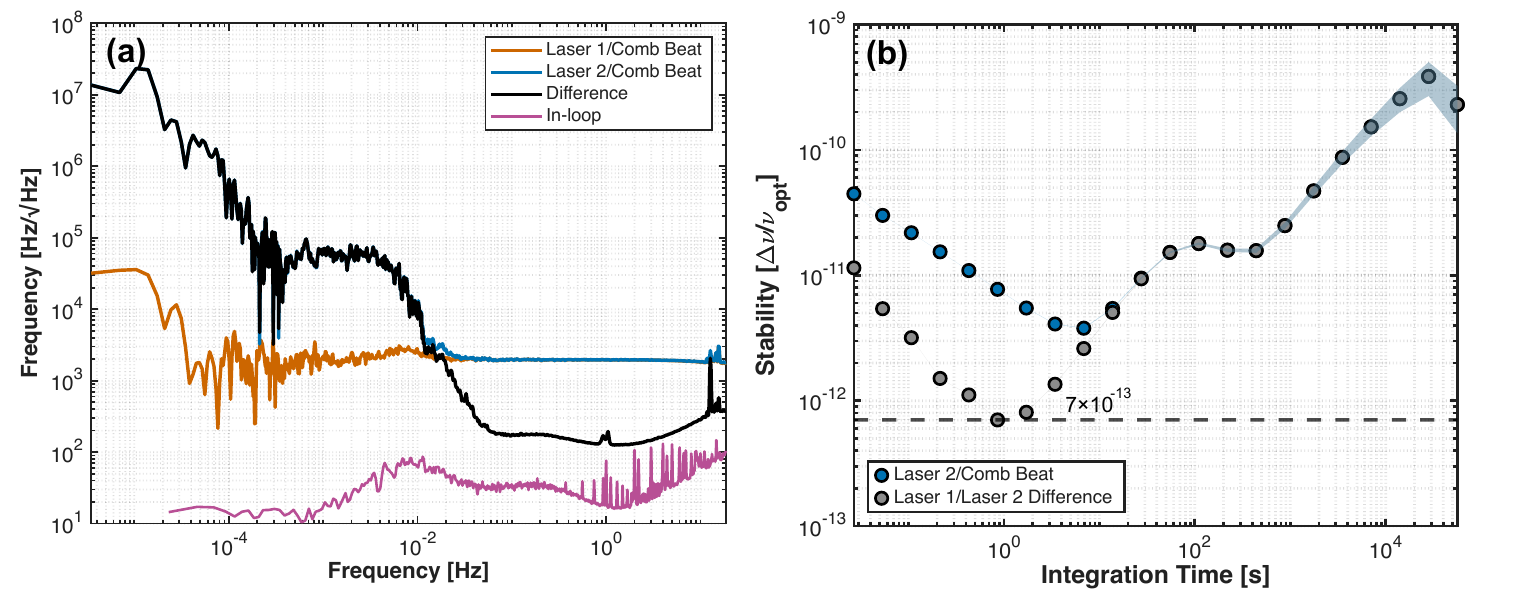}
    \caption{(a) Frequency-noise amplitude spectral density of the out-of-loop beat note frequency measurement of each laser with the optical frequency comb. Laser 1 is acting as the reference laser, which is frequency locked to the comb. Trace 2 measures the stability of laser 2 locked to laser 1 with respect to a different comb line. The differential provides an out-of-loop measure of the transfer stability. Low frequency performance is limited by the modulation asymmetries (0.3 mHz - 40 mHz), and residual interferometer length noise ($< 0.3$ mHz). We hypothesize the relative stability measurement is limited by the common-mode cancellation of comb frequency noise by the two out-of-loop phasemeters. A separate 12 hour in-loop (error signal) measurement, with control system noise suppressed shows than a factor of $10$ clearance. (b) Allan deviation of the measurement showing stability of $7\times10^{-12}$ at 1 second measured out-of-loop with respect to the frequency comb, and $7\times10^{-13}$ at 1 second relative stability between the two lasers, normalized to the laser 2 optical frequency.}
    \label{fig:psd}
\end{figure}

The noise spectrum of both comb beat notes was computed using a log-frequency power spectral density in order to cover the full frequency span of the measurement \cite{trbs2006improved-d07}. The resultant frequency-noise amplitude spectral density is shown in Fig \ref{fig:psd} (a) alongside the Allan deviation in Fig. \ref{fig:psd} (b).

In the frequency-noise amplitude spectral density we have four traces. Traces 1 and 2 are the respective out-of-loop beat frequency measurements of each laser with the optical frequency comb. Trace 3 is the difference between trace 1 and 2, and we see suppression of common frequency noise between the two measurements in the high Fourier frequency band (above 40 mHz) by approximately a factor of 10. At these frequencies, we hypothesize to the limited by common-mode noise cancellation of the optical frequency comb from the two independent Moku:Pro phasemeters. We also include trace 4, an in-loop differential phase measurement of the error signal computed without the small angle approximation, which lies an order of magnitude below the out-of-loop differential. The in-loop differential phase amplitude spectral density is converted to frequency noise through interferometer FSR. We note, however, that the in-loop measurement was not synchronous with the 3.5 day measurement, and was carried out the day after. Furthermore, control system imposed noise is suppressed in the in-loop measurement and can thus not be excluded in the in-loop measurement. A measurement of the Moku:Pro phasemeter self-noise was also measured to be below 1 mHz/$\sqrt{\text{Hz}}$ for stationary signals over the measurement band.

Below 40 mHz, we observe a cyclic noise feature which clamps at $5\times10^{4}$ Hz/$\sqrt{\text{Hz}}$. This noise feature is driven by modulation asymmetries which result in ellipticity of the resultant DEHoI measurements. The removal of these artifacts has been demonstrated both by up-conversion through phase modulation injection \cite{Ngo2016SuppressingInterferometers}, and via real-time constellation correction processes \cite{sambridge2025dual-quadrature-b90,Hu2015}. In Fig. \ref{fig:psd} (b), we see that the relative stability reaches a minimum of $7\times10^{-13}$ at 1 second as measured by the differential between the two out-of-loop measurements. There also exists a prominent hump in the Allan deviation at 100 seconds, corresponding with the cyclic noise process discussed above. 

At and below 0.3 mHz in Fig. \ref{fig:psd} (a) we also see the residual coupling of interferometer length noise into the measurement, which continues to be dominant over longer timescales in (b). We further characterize the residual length noise coupling in the following section.

\section{Cross-coupling Measurements}
As detailed in the publication \cite{luiten2020}, the common-mode cancellation of interferometer length noise is dependent on the wavelength separation between the lasers under test. As described in Eqn. 4 of \cite{luiten2020}, the scaling between the phase measurements of different wavelengths is given by $\nu_1n_1/\nu_2n_2$. The residual length coupling in the absence of this correction for small wavelength separations can then be given by Eqn. \ref{eq:residualcoupling}:
\begin{align}
    \Delta\nu_{L}(t) &= \left(1 - \frac{\nu_1n_1}{\nu_2n_2}\right)\frac{\Delta L(t)\nu_2 n_2}{L_{\text{IF}}},
    \label{eq:residualcoupling}
\end{align}
where $\Delta\nu_{L}(t)$ is the frequency error due to residual length coupling. It is dependent on the frequencies of the two lasers under test, $\nu_1$ and $\nu_2$, and their effective refractive index in the waveguide ($n_1$ and $n_2$). It is then proportional to the dynamic length change $\Delta L(t)$ divided by the length of the interferometer. For the two lasers under test in this demonstration, we compute a residual coupling of 443 ppm, from the lasers' internal wavelength reading. 

For our wavelength spacing between the two lasers, we can also directly measure the coupling by locking both lasers to the optical frequency comb. By then measuring the frequency stabilization readouts for each interferometer and their differential, the length noise can be characterized. 

\begin{figure}
    \centering
    \includegraphics[width=0.95\linewidth]{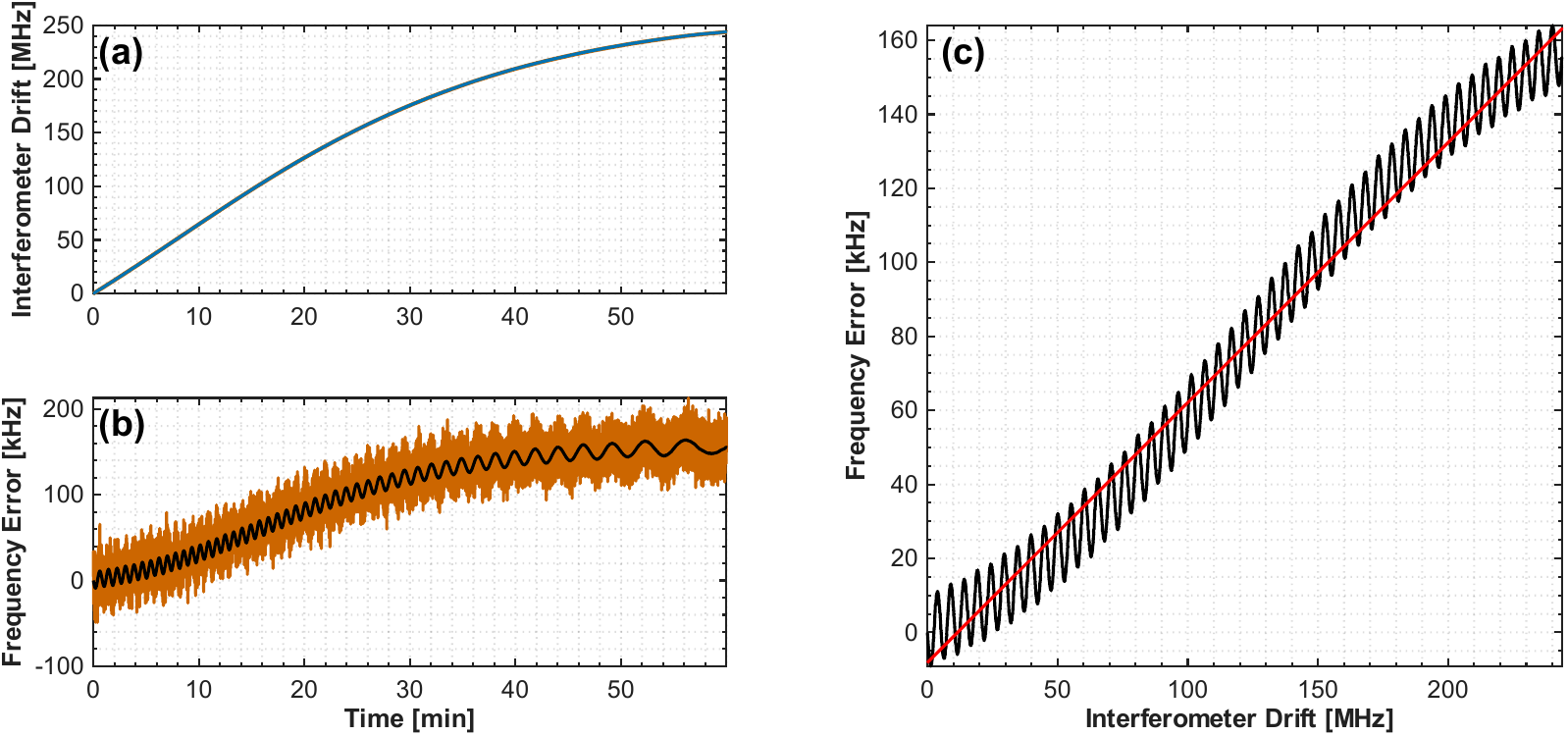}
    \caption{Plot (a) shows the measured interferometer drift using laser 1 and laser 2 (visually identical), calibrated to equivalent frequency change. Plot (b) shows the free-running frequency stabilization error signal, with a 500 ms moving averaged trace (black) to highlight cyclic error coupling. Plot (c) shows the cross-coupling relation between interferometer drift and residual coupling into the frequency stabilization readout. The first order linear fit (red) indicates a 700 ppm coupling coefficient of residual length noise.}
    \label{fig:pathlengthnoise}
\end{figure}

To do this, we record a 1 hour measurement with both lasers offset frequency locked at their respective wavelengths to the nearest neighbor comb line. The frequency stabilization system, while disengaged, then records the interferometer phase drift as seen by both lasers, corresponding to two wavelength separated measurements of the interferometer length noise. The frequency stabilization error signal is also recorded. As both lasers are comb stabilized for this measurement, this error signal should remain at a fixed DC value in the absence of any residual interferometer length coupling. In Fig. \ref{fig:pathlengthnoise} we plot the results of this measurement, with (a) showing the individual (Laser 1) interferometer drift, calibrated to frequency using the known free-spectral range of the interferometer. The measurement from Laser 2 is visually identical on this scale. 

Figure \ref{fig:pathlengthnoise} (b) shows the free-running frequency stabilization error signal, with the DC offset baselined to the start of the measurement. We also show a 500 ms moving averaged readout in black on the same plot to highlight the cyclic error present in the readout. This residual error arises from modulation asymmetries and modulation-quadrature cross coupling, resulting in a parasitic phasor about the main measurement. This results in a noise process with a spectral shape proportional to the cosine of the interferometer phase dynamics. As laser frequency drift increases at low Fourier frequencies, this results in the characteristic noise roll-up and saturation feature observed from 40 mHz to 0.3 mHz. The specific frequency at which this occurs is therefore laser specific, with larger dynamics leading to an onset at higher Fourier frequencies. The lower bound of this band is set by the crossover with residual interferometer length coupling. The cyclic error saturates once the interferometer phase sweeps through multiple fringes, whereas the length coupling of Eqn. \ref{eq:residualcoupling} continues to grow with the thermal drift of the interferometer as the Fourier frequency decreases. For cyclic error, previous work has demonstrated that driving the interferometer phase dynamics at a well-determined frequency is able to upconvert this noise process to the drive Fourier frequency \cite{Ngo2016SuppressingInterferometers}, which can then be filtered from the measurement and control system. This removes the noise process from the measurement band of interest.

The cross-coupling between the interferometer length and the frequency stabilization error signal is then shown in the parametric plot in Fig. \ref{fig:pathlengthnoise} (c). Here we see a clear first order response with a coupling coefficient (slope) of $7\times10^{-4}$ (700 ppm). This is within a factor of two of the predicted value, and probable causes of the discrepancy are likely offsets in the lasers' internal wavelength reading, and polarization-state differences between the two wavelengths. 

While the residual coupling could in principle be reduced by applying the scaling correction, $\nu_1n_1/\nu_2n_2$, prior to forming the differential error signal, the current architecture operates in the phasor domain and does not perform explicit per-channel phase extraction. Implementing this correction requires per-channel arctangent and phase unwrapping operations, representing a non-trivial modification to the FPGA control loop to avoid quantisation errors given the comparatively small wavelength separation relative to prior demonstrations \cite{luiten2020}. Thus, this improvement to the architecture is identified as a near-term improvement to extend long term stability.

\section{Tunability}
As a final demonstration of the flexibility of this platform we show software lock-point tunability. Here Laser 1 remains referenced to the optical frequency comb while Laser 2 is locked and modulated while stabilized, achieved via digital modulation of the frequency reference demodulation angle. This implementation enables the signal injection to absolve the requirement for any further calibration, both in terms of the modulation depth, which is proportional to the interferometer FSR, and the modulation frequency, which is directly related to the digital synthesis sampling rate. 

\begin{figure}[h!]
    \centering
    \includegraphics[width=0.75\linewidth]{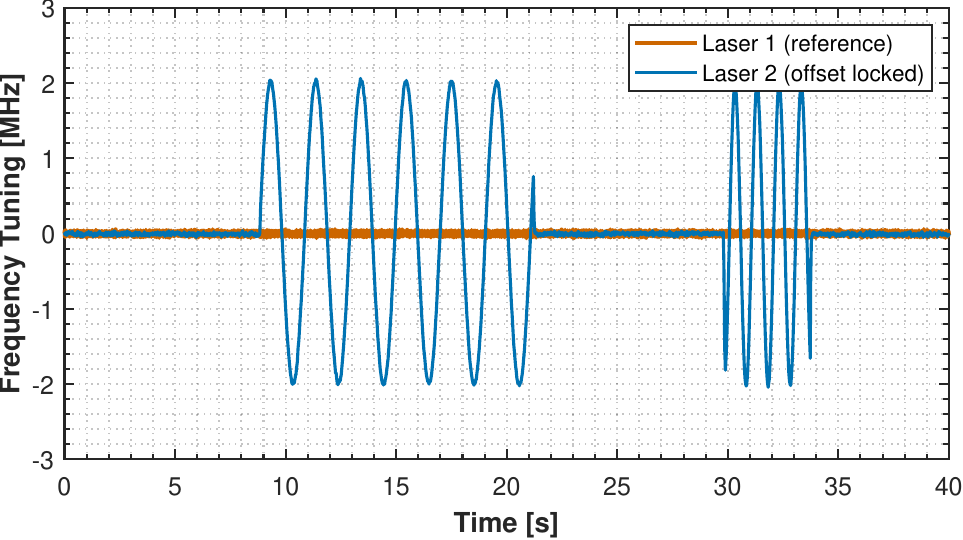}
    \caption{Demonstration of dynamic tuning while retaining lock-point. Here laser 1 maintains lock to the external optical frequency comb. Laser 2 is offset locked to laser 1. A sinusoidal tuning injection at 500 mHz with 2 MHz depth is turned on at 9 seconds and switched off at 21 seconds. The modulation frequency is then increased to 1 Hz and injected from 30 seconds to 34 seconds. After both injections, the system is able to recover to the original lock point.}
    \label{fig:tune}
\end{figure}

In the demonstration shown in Fig. \ref{fig:tune}, we switch the frequency tuning on at a depth of 2 MHz and tuning frequency of 500 mHz. The system is then able to recover to the original lock point when the modulation is switched off mid cycle. The frequency is then increased to 1 Hz for a second interval before returning to the original lock point again.

While we are demonstrating modulation here, this method can also be used to continuously tune and set the lock point of the system at arbitrary offsets, with resolution limited by the 16-bit word used to address the demodulation angle. The resolution in frequency can then be determined by dividing the interferometer FSR (8 MHz) by the address word length ($2^{16}$). This gives a frequency tuning resolution of 122 Hz.

\section{Scalability}
The architecture we demonstrate here is extensible to interrogate multiple lasers, beyond the two of this initial demonstration. To understand the implications of scaling, we first consider the additional hardware requirements. For each laser under test we require a phase modulator (EOM in our case), one RF DAC channel on the FPGA to provide the digital interferometric code and an input coupler port on the $1\times N$ input combiner. If feedback is to be implemented, an additional low-speed DAC is required. Each of these scale as $N$ with the number of lasers. For FPGA resource usage, previous double-encoded digital interferometry demonstrations have shown 30 simultaneous channels scaling comfortably on legacy Kintex-7 Xilinx FPGA platforms \cite{chan2025narrow-band-658}, and this may be separable over multiple discrete, time synchronised systems.

At the photodetector, the DC power is given by the sum of the total power of all fields. The saturation limit of the detector is therefore scaling per Eqn. \ref{eq:DCpower}
\begin{align}
    P_{DC} = \sum_{i = 1}^N \alpha_i P_i/2N,
    \label{eq:DCpower}
\end{align}
at each photodiode, where $P_i$ is the input power of each laser prior to the input combiner and $\alpha_i$ is the loss for signal $i$ through the interferometer itself. The scaling by $N$ is due to the loss through the input combiner.

The RF power carries all the demodulation within the detector bandwidth and thus the balanced detector will have a fixed RF saturation point which limits the maximum power. To calculate the RF power limit, we assume the same mode of operation as our current implementation, with each beat note between the lasers located outside the bandwidth of the photodetector. Given this regime, the root-mean-square RF power decreases as given by Eqn. \ref{eq:RFpower}:
\begin{align}
    P_{RF} = \sqrt{ \sum^N_{i=1} (\alpha_i P_i/2N)^2}.
    \label{eq:RFpower}
\end{align}
We see that for equal power, Eqn. \ref{eq:RFpower} simplifies to a $1/\sqrt{N}$ dependence on the number of lasers, assuming the interferometric loss and the input power of each laser remain fixed during scaling. Depending on the full-scale range of the photodetector ADC, quantization limits follow an analogous scaling with number of lasers.

Considering solely the code, it itself is able to scale in principle up to the length of the generating m-sequence, which in our instance has 2047 unique delays. In our implementation we have used a delay of 60 symbols between the two channels, which then provides up to 34 channels of room within the existing modulation framework. This is however a conservative estimate, and could be increased further by dovetailing the measurement delays of different channels. We note that increasing the length of the code is possible, at the loss of measurement bandwidth.

The second code related consideration is the code cross-correlation. Our implementation has the laser carrier frequency separation greater than the photodetection bandwidth and spread-spectrum bandwidth of the code. This means that all heterodyne cross-correlation terms between lasers are not within photodetection range. As a result the code cross-correlation is between the self-homodyne terms of each code, as described in Eqn. \ref{eq:pd_dual}. For an m-sequence pseudo-random code, the theoretical suppression scales as $1/L_c$ in amplitude for each uncorrelated signal, where $L_c$ is the number of symbols in the code. For the dual m-sequence QPSK code used here, the suppression is $2/L_c$, as the real-valued photodetector signal contributes two $1/L_c$ residuals to the complex demodulated output. Following Sibley et al.~\cite{sibley2020crosstalk-80c}, each crosstalk residual scales with the ratio of the signal amplitudes, which are proportional to optical power per channel at the photodetector, given by $\alpha_i P_i$. Assuming uncorrelated signals, these add in quadrature yielding Eqn. \ref{eq:codesuppression}:
\begin{align}
\text{Crosstalk} &= \frac{2}{L_c}\frac{\sqrt{\sum_{j=1, j\neq i}^{N} \alpha_j^2P_j^2}}{\alpha_iP_i},
\label{eq:codesuppression}
\end{align}
which describes the crosstalk coupling into the $i$th channel from the quadrature sum of all other channels. It scales with the power of each additional signal on the photodetector. When laser power and interferometer losses are equal for each signal, this reduces to a scaling of $\sqrt{N - 1}$. 

The shot noise limit for scaling digital interferometric systems can be calculated in the same way as presented by Sibley \cite{sibley2020crosstalk-80c}, but refactoring for the homodyne terms being the signal of interest. This leads to a phase shot noise level that scales with the number of lasers as per Eqn. \ref{eq:shotnoise}
\begin{align}
\phi_{\text{Shot},i}[\text{cycles/}\sqrt{\text{Hz}}] &= \sqrt{\frac{\hbar c}{2\pi}\frac{\sum_{i=1}^NP_{i,PD}}{\rho\lambda_i P_{i,PD}^2}},
    \label{eq:shotnoise}
\end{align}
where $\lambda_i$ is the wavelength of the laser being measured, $\rho$ is the detector quantum efficiency, and $P_{i,PD}$ is the power of the corresponding laser incident at the photodiodes. In the case where all lasers are power balanced, and we substitute in the power at the input combiner for each laser, $P_i$, using the form used in Eqn. \ref{eq:DCpower}, this reduces Eqn. \ref{eq:shotnoise} to Eqn. \ref{eq:shotnoise_equal}:
\begin{align}
\phi_{\text{Shot},i}[\text{cycles/}\sqrt{\text{Hz}}] &= \sqrt{\frac{\hbar c N^2}{\pi\rho\lambda_i \alpha_i P_i}}.
    \label{eq:shotnoise_equal}
\end{align}
In addition to this are the control loop, laser frequency noise and differential phasemeter common-mode rejection characteristics of a given implementation. These will factor into the final transfer performance and are likely to be the dominant limiting process for systems with a small number of lasers. 

\section{Conclusion}
We have demonstrated frequency stabilization between two fiber lasers via a passive path imbalanced Mach Zehnder interferometer as a shared fiber frequency reference, achieving an upper bound of sub-kHz/$\sqrt{\text{Hz}}$ stability at Fourier frequencies above 40 mHz. We also demonstrate a fine resolution lock-point tunability which shows the flexibility of this system. The architecture offers a cost effective, scalable and readily deployable solution compared to the use of optical frequency combs for stabilizing multiple lasers at separated frequencies. Further we also propose a way forward to improve stability at Fourier frequencies below 40 mHz, by implementing cyclic noise upconversion and minimizing residual length coupling. The demonstrated relative stability and software-controlled tunability are relevant to the frequency-matching requirements of TF-QKD and CV-QKD, and to two-color applications such as tunable microwave generation. Extension to remote laser sources with link stabilization, and to the larger wavelength separations required for terahertz synthesis, present avenues for future work.

\begin{acknowledgments}
HK and VBA would like to acknowledge the support from the Swedish Research Council (VR starting grant 2023-0519), G{\"o}ran Gustafsson Prize for early career researchers and the Wallenberg Center for Quantum Technology (WACQT) in Sweden. RLW and CPB are affiliated with the OzGrav Australian Research Council Centre of Excellence. CPB is additionally affiliated with Forge Photonics PTY LTD and COMBS Australian Research Council Centre of Excellence.\\

The authors declare no competing interests.
\end{acknowledgments}


%

\end{document}